\documentstyle[12pt,epsf]{article}

\newcommand {\be}  {\begin{equation}}
\newcommand {\ee}  {\end{equation}}
\newcommand {\bea} {\begin{eqnarray} \nonumber }
\newcommand {\eea} {\end{eqnarray}}

\begin{document}

\title{Critical Behavior of the 4D Spin Glass in Magnetic Field} 

\author{Enzo Marinari, Carla Naitza and Francesco Zuliani\\[0.5em]
{\small  Dipartimento di Fisica and INFN, Universit\`a di Cagliari}\\
{\small  Via Ospedale 72, 09100 Cagliari (Italy)}\\[0.3em]
{\small   \tt marinari, naitza, zuliani@ca.infn.it}\\[0.5em]
}

\date{February 14, 1998}

\maketitle

\begin{abstract}
We present numerical simulations of the 4D Edwards Anderson Ising spin 
glass with binary couplings.  Our results, in the midst of strong 
finite size effects, suggest the existence of a spin glass phase 
transition.  We present a preliminar determination of critical 
exponents.  We discuss spin glass susceptibilities, cumulants of the 
overlap and energy overlap probability distributions, finite size 
effects, and the behavior of the disorder dependent and of the integrated 
probability distribution.
\end{abstract}  

\vfill

\begin{flushright}
cond-mat/9802224
\end{flushright}

\thispagestyle{empty}
\newpage

\section{Introduction}

The Edwards-Anderson model of spin glasses \cite{EDWAND} turns out, 
not astonishingly, to be hard to be understood.  More surprisingly (at 
the start) also the Sherrington Kirkpatrick mean field version of the 
model \cite{SK} appears to be very complex.  Many unprecedented features 
appear: for example the (spin glass) phase transition survives the 
presence of a finite magnetic field \cite{DEATHO} under the AT line.  
The Parisi Replica Symmetry Breaking (RSB) solution \cite{PARISI} 
appears to describe accurately the model in the low $T$ broken phase 
(and it is believed to be the real solution of the model).

The relevant question is now trying to establish how many of the 
features of the Parisi solution survive when discussing finite 
dimensional spin glass models.  For example the presence of a phase 
transition in a finite magnetic field, that is implied by the Parisi 
solution \cite{PARISI}, is not compatible with the point of view of 
the {\em droplet model} \cite{DROPLET}, where one expects the 
transition to be removed from the action of a small magnetic field.

As usual in a complex theoretical scenario numerical simulations try 
to play a role (for a review of recent simulations of spin glass 
models see \cite{OUR-REVIEW}). Here we will report numerical 
simulations that thermalize large lattices with large number of 
samples: we will try to show many signatures hinting about the 
presence of a phase transition, and we will be plagued by large 
finite size effects.

Numerical simulations of spin glasses with non-zero magnetic field 
have now a long story. Maybe the first work on the subject is the one 
by Sourlas \cite{SOURLAS-A,SOURLAS-B}. The numerical simulations 
of \cite{CPPS-A,CPPS-B} (see also the criticism in \cite{HUSFIS} and 
the reply \cite{CPPS-C}) use the concept of energy overlap, that turns 
out to be very relevant in this study: these simulations were giving a 
first evidence, but the computers of the time allowed too small 
lattices for getting conclusive results. Simulations of \cite{GRAHET} 
measure constant susceptibility curves \cite{SINHUS}, 
and detect the existence of an  AT line. On the contrary the work of 
\cite{KAWITO} does not detect one, while \cite{BCPPRR,CPRR,PICRIT-A} 
are more on the yes side. Again, the work of \cite{ANMASV} is negative 
on the existence of a transition, while \cite{PICRIT-B} claims the 
difficulty of establishing a firm conclusion (maybe a wise approach).
At last recent numerical work using a dynamic approach has been able 
to claim, again, the existence of a clear transition to a spin glass 
phase \cite{MAPAZU,PARIRU}.

Here, as we will explain in the next sections, we will find evidence 
that we consider strongly suggestive of the existence of a phase 
transition, and we will present very preliminary determinations of 
critical exponents.  We will also find that even on large lattices (on 
current standards) finite size effects are dramatically strong, and 
we will discuss them in detail.

\section{Numerical Simulations}

We will discuss here the $4D$ Edwards-Anderson spin glass system 
\cite{EDWAND} with bimodal quenched random couplings $J=\pm 1$.  Our 
numerical simulations have been using the {\em parallel tempering} 
approach \cite{TEMPERING}, that makes a real difference in the 
simulations of systems with quenched disorder. The interested reader 
will find for example in the last of \cite{TEMPERING} an introduction 
to optimized Monte Carlo methods (including tempering and, more in 
general, the multi-canonical approaches).

\begin{table}
\centering
\begin{tabular}{|c||c|c|c|c|c|c|c|} \hline
$L$ & Thermalization & Equilibrium & Samples & $N_{\beta}$ & $\delta T$ &  
$T_{min}$& $T_{max}$\\ \hline \hline
3   &   20000  &  20000    & 2560   &  19  &  0.1  & 1.0 & 2.8\\ \hline
5   &   80000  &  80000    & 1920   &  19  &  0.1  & 1.0 & 2.8\\ \hline
7   &  200000  & 200000    &  960   &  46  &  0.04 & 1.0 & 2.8\\ \hline
9   &  150000  & 150000    &   64   &  66  &  0.04 & 1.0 & 3.6\\ \hline
\end{tabular}
\caption[0]{Parameters of the tempered Monte Carlo runs.
\protect\label{T-PARAME}}
\end{table}

We report in table (\ref{T-PARAME}) the parameters relevant for our 
tempered simulations: for each lattice size we give the number of 
discarded thermalization sweeps and of the sweeps used for 
measurements, the total number of samples that we have analyzed, the 
number of temperatures used in each tempered run (i.e. the number of 
copies of the system updated in parallel at different $\beta$ values 
and among which the temperature values have been swapped), the 
temperature increment, the minimum and the maximum temperature.

The $\beta$ values have been chosen, as customary and reasonable, in order 
to keep the acceptance factor of the tempering $\beta$ swap of order 
$\frac12$. 
In our case we have a $\beta$ acceptance ratio close to $0.8$ for $L=3$ that 
goes down to a number close to $0.6$ at $L=9$.

Our runs are surely well thermalized for $L\le 7$: for each copy 
$\beta$ has visited all possible values at least a few times, and the 
system has never been stuck. $L=9$ is more delicate, and we are not 
sure that the data points with lower $T$ values are fully thermalized: 
we have repeated different trial runs, with different values of the 
parameters, and the one we report here are the ones that turn out to 
be better thermalized. Difference among the different runs were in any 
case minor, and  a possible remanence of on thermalization effects 
would affect only minor issues that we will point out in the following. 
Here we will only insist on features that are surely representing 
thermal equilibrium even on the $L=9$ lattice.

\section{Spin Glass Susceptibilities}

In this section we will discuss the overlap and the energy overlap 
susceptibilities. We will show the signature of a spin glass 
like phase transition. We will discuss the location of the critical 
temperature and the determination of the critical exponents.

We consider two real replicas of the system with spins $\sigma_{i}$ and 
$\tau_{i}$, and the local energy operator 

\be
  \epsilon_{i}^{(\sigma)}\equiv \frac{1}{2D}
  \sigma_{i}\sum_{j}J_{i,j}\sigma_{j} \ ,
\ee
where the sum runs over first neighbors of the site $i$ (on a $4D$ 
hypercubic lattice of linear size $L$ and volume $V=L^{4}$). The {\em 
overlap} operator is defined as

\be
  q\equiv\frac{1}{V} \sum_{i}\sigma_{i}\tau_{i}\ ,
\ee
and the {\em energy overlap} operator as

\be
  q_{E} \equiv \frac{1}{V} 
  \sum_{i}\epsilon^{(\sigma)}_{i}\epsilon^{(\tau)}_{i}\ .
\ee
This operator plays a crucial role, since it allows to distinguish a 
possible trivial replica symmetry breaking from a non-trivial 
breaking. In the Mean Field RSB $q_{E}$ and $q^{2}$ coincide, while a 
non-trivial behavior of $q$ induced by the presence of interfaces 
would generate a trivial $q_{E}$. Detecting a non-trivial behavior 
of $q_{E}$ is strong evidence for a RSB like behavior. We will 
consider the probability distribution of the overlap for a given 
sample of the quenched disorder,  $P_{J}(q)$, and the same for the 
energy overlap, $P^{E}_{J}(q)$. We will call $P(q)$ and $P^{E}(q)$ the 
probability distribution integrated over the quenched disorder.

We define the {\em overlap susceptibility} as

\be
  \chi_{q}\equiv V \left( E(q^{2})-E(q)^{2}  \right)\ ,
\ee
where by $E(\ \cdot\ )$ we denote the combined operation of a thermal 
average and an average over the quenched disorder $J$ (in the usual 
notation $E(\ \cdot\ )=\overline{\langle\ \cdot\ \rangle}$). We 
define the {\em energy overlap susceptibility} as 

\begin{figure}
\begin{center}
\leavevmode
\epsfysize=250pt
\epsffile{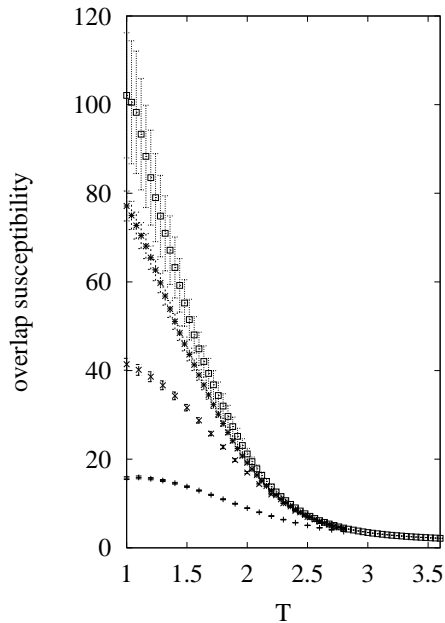}
\end{center}
\caption[0]{\protect\label{F-CHI}
$\chi$ as a function of $T$. Empty squares 
for $L=9$, asterisks for $L=7$, crosses for $L=5$ and horizontal bars 
for $L=3$.}
\end{figure}
\begin{figure}
\begin{center}
\leavevmode
\epsfysize=250pt
\epsffile{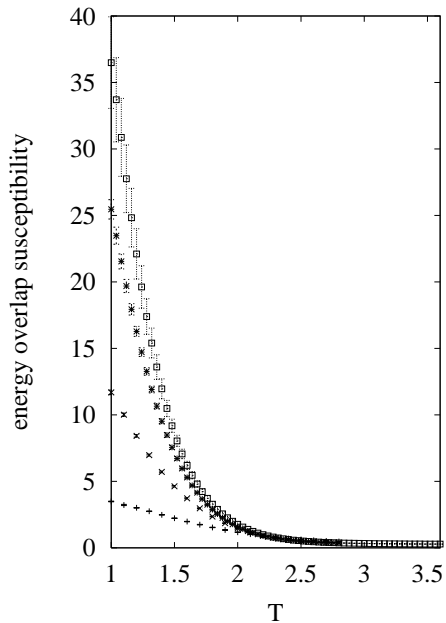}
\end{center}
\caption[0]{\protect\label{F-CHI-E}
  As in figure (\ref{F-CHI}) but for $\chi_{q_{E}}$.}
\end{figure}


\be
  \chi_{q_{E}}\equiv V \left( E(q_{E}^{2})-E(q_{E})^{2}  \right)\ .
\ee
In figure (\ref{F-CHI}) we plot the overlap susceptibility $\chi$ as a 
function of $T$ for $L=3$, $5$, $7$ and $9$, and in figure 
(\ref{F-CHI-E}) we plot the energy overlap susceptibility $\chi_{E}$.
The statistical errors are computed using a jack-knife algorithm.

Both susceptibilities show a divergence in the low $T$ region. 
The comparison of figure (\ref{F-CHI}) with the analogous plot of 
reference \cite{PICRIT-B} shows a difference: we do not see a cusp 
but a clear divergence (not only for $L=7$ but also for $L=9$ down to 
$T=1.0$). We believe this is very well explained from some non 
complete thermalization for the larger lattices of reference 
\cite{PICRIT-B} (this possibility is proposed and discussed in 
reference \cite{PICRIT-B} itself: a non complete thermalization
has exactly the effect of smoothing the divergence, since correlations 
on large scale cannot be created): these measurements are very 
delicate. As we have already discussed we are completely safe as far 
as the $L=7$ lattice is concerned, but $L=9$ was clearly our limit.
It is important to stress that this is the only point of discrepancy 
with reference \cite{PICRIT-B}: as far as other (less sensitive) 
quantities are concerned (and even for the susceptibility on the two 
smaller lattice sizes) we find exactly the same results. Reference 
\cite{PICRIT-B} does not analyze the quantities based on the energy 
overlap.

As we have said the divergence of the two susceptibilities is clear.  
We are able to follow the divergence on thermalized lattices down to 
$T=1.0$.  The fact that also the energy overlap susceptibility 
diverges makes a stronger case for a RSB like spin glass phase.  The 
analysis of reference \cite{MAPAZU} hints that at $h=0.4$ is close or 
slightly lower than $1.5$.  We just note at this point that the data 
of figures (\ref{F-CHI}) and (\ref{F-CHI-E}) are fully compatible 
with this value.

\begin{figure}
\begin{center}
\leavevmode
\epsfysize=250pt
\epsffile{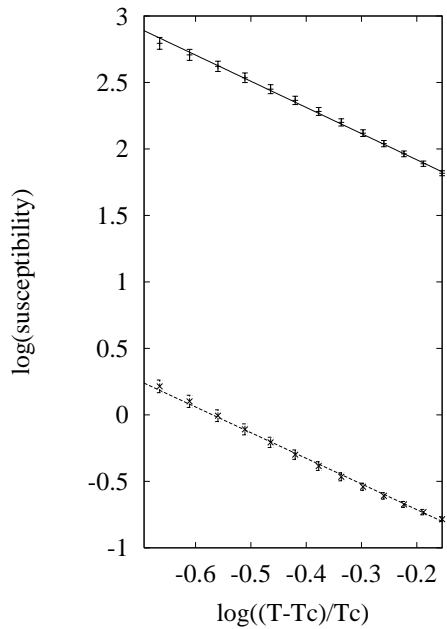}
\end{center}
\caption[0]{\protect\label{F-GAMMA}
  $\log(\chi_{q})$ and best fit to a power law (continuous line) and 
  $\log(\chi_{q_{E}})$ and best fit to a power law (dashed line) 
  versus the logarithm of the reduced temperature.}
\end{figure}

One can see in figures  (\ref{F-CHI}) and (\ref{F-CHI-E}) that there 
is a temperature region where the behavior is asymptotic (i.e. 
measurements on the $L=9$ lattice are compatible with the ones taken on 
smaller lattices). We use this region to fit the susceptibility as a 
function of the  reduced temperature $\frac{T}{T-T_{c}}$, with 

\bea
  \chi_{q}(t)     &\simeq& A_{q}t^{-\gamma}\ , \\
  \chi_{q_{E}}(t) &\simeq& A_{q_{E}}t^{-\gamma_{\epsilon}}\ .
  \protect\label{E-GAMMA}
\eea
We will use the value $T_{c}=1.4$ (in agreement with the results of 
\cite{MAPAZU} and with evidence that will be discussed later in this 
note).  The fits in the region of $T$ going from $2.1$ to $2.6$ (that 
we show in figure (\ref{F-GAMMA})) give $\gamma=1.97$ and 
$\gamma_{\epsilon}=1.93$.  $\chi_{q}$ has a stronger $L$ dependence 
than $\chi_{q_{E}}$.  We plot in figure (\ref{F-GAMMA}) the logarithm 
of $\chi_{q}$ and $\chi_{q_{E}}$ versus $\log(t)$, and the best fits 
to the form (\ref{E-GAMMA}).

The conclusion of this analysis is that the numerical data are 
compatible with a divergence at $T_{c}\simeq 1.4$ with an exponents
$\gamma$ and $\gamma_{\epsilon}$ close to $2$. 

We have also carried through a standard finite size scaling analysis,
by selecting $T_c$ and  the critical exponents in such a way to have curves
at different lattice size collapsing together as well as possible. As usual for
not very accurate data (as is unfortunately frequently the case for
numerical simulations of complex systems)
this analysis does not give unambiguous results, but only hints
reasonable and preferred set of values. 
We use for $\chi_q$ the leading scaling form

\be
  \chi_q = L^{\frac{\gamma}{\nu}} \overline{\chi}_q
  \left( L^{\frac{1}{\nu}}(T-T_c)\right)\ ,
\ee
and the same form for $\chi_{q_E}$. By looking at $\chi_q$ we find
that $T_c=1.4$, $\frac{1}{\nu}=0.7$ and $\frac{\gamma}{\nu}=1.3$ give
a very good fit. The same values (with $\gamma_\epsilon=\gamma$) also
give a very good scaling behavior for $\chi_{q_E}$.

If one would take a higher value of $T_{c}$ (that could be suggested 
by a possible interpretation of the crossover regime we will discuss 
in the next section, but is discouraged by the dynamical data of 
\cite{MAPAZU}), for example $T_{c}=1.8$, we would find a higher value 
of $\nu$, of the order of two, and $\frac{\gamma}{\nu}\simeq 1$. 
Again, the finite size scaling analysis and the study of the 
asymptotic $T$ dependence would be consistent at this effect.

\section{Skewness and Kurtosis}

In order to qualify the probability distribution of the overlap and of 
the energy overlap we will define and analyze their kurtosis (the 
Binder cumulant) and their skewness.  For zero-field deterministic and 
disordered statistical systems the Binder cumulant $g$ of the order 
parameter is a very good signature of the phase transition: curves of 
$g$ versus $T$ cross at the critical point, since the kurtosis of the 
probability distribution of the order parameter at the critical point 
is an universal quantity.  In the infinite volume limit $g$ goes to 
zero in the warm phase: it goes to one in a ferromagnetic phase, and 
to a non-trivial function in the broken phase of the Parisi RSB 
solution of the mean field spin glass theory.

\begin{figure}
\begin{center}
\leavevmode
\epsfysize=250pt
\epsffile{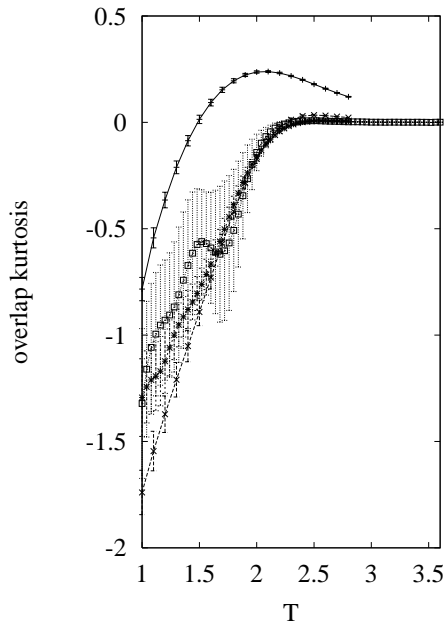}
\end{center}
\caption[0]{\protect\label{F-KUR}
  As in figure (\ref{F-CHI}) but for the overlap kurtosis $g$.}
\end{figure}

Here, since we have a non zero magnetic field, we have to consider 
connected expectation values. We define the overlap kurtosis on a 
lattice of linear size $L$ as

\be
  g(T) \equiv \frac32-\frac12 
  \frac{ E\left(\left(q-E(q)\right)^{4}\right) }
       { E\left(\left(q-E(q)\right)^{2}\right)^{2} }
  \equiv  \frac32-\frac12 \frac{\chi_{q}^{(2)}}{\chi_{q}^{2}}\ ,
  \protect\label{E-KUR}
\ee
which defines $\chi^{(2)}_{q}$. We define the kurtosis for the energy 
overlap $g_{\epsilon}$ in the analogous way, by using $q_{E}$. We 
define the skewness of the probability distribution of the overlap as

\be
  s(T) \equiv 
  \frac{ E\left(\left(q-E(q)\right)^{3}\right) }
       { E\left(\left(q-E(q)\right)^{2}\right)^{\frac32} }
  \equiv  \frac{\chi_{q}^{(\frac32)}}{\chi_{q}^{\frac32}}\ ,
  \protect\label{E-SKE}
\ee
which defines $\chi^{(\frac32)}_{q}$. In the definition of the skewness 
of the energy overlap probability distribution $s_{\epsilon}$ one 
substitutes  $q_{E}$ to $q$. We note now that the study of the 
$q_E$ probability distribution turns out to be very important.
We plot $g(T)$ for the four different lattice volumes in figure 
(\ref{F-KUR}). In this and in next plots errors are from 
sample-to-sample fluctuations evaluated with a jack-knife analysis.

The difference with the usual zero field picture is strong.  There is 
a clear change of regime close to $T=2$ (the critical point at $h=0$).  
The $L=3$ system is small and different, and never really gaussian in 
our $T$ range.  For $L\ge 5$ $g$ becomes non-trivial when $T$ becomes 
smaller than $2$.  The $q$-kurtosis in this region does not change 
much with size.  In the statistical error the values of the $L=5$, $7$ 
and $9$ lattice are compatible (but maybe at $T\simeq 1$ where also 
small non-equilibrium effects have to be accounted for).  The fact 
that the kurtosis does not depend on $L$ and is non-trivial is very 
clear from our data.  Again, the traditional crossing behavior is 
completely absent (likely even in the infinite volume limit) in our 
data.  Two things must be stressed here: first of all that the existence 
or non-existence of a crossing also depends from the shape of the critical 
asymptotic probability distribution, and second that we cannot 
exclude, and on the contrary we clearly detect (see for example next 
figure, (\ref{F-KUR-E})) the existence of very strong finite size 
effects.

\begin{figure}
\begin{center}
\leavevmode
\epsfysize=250pt
\epsffile{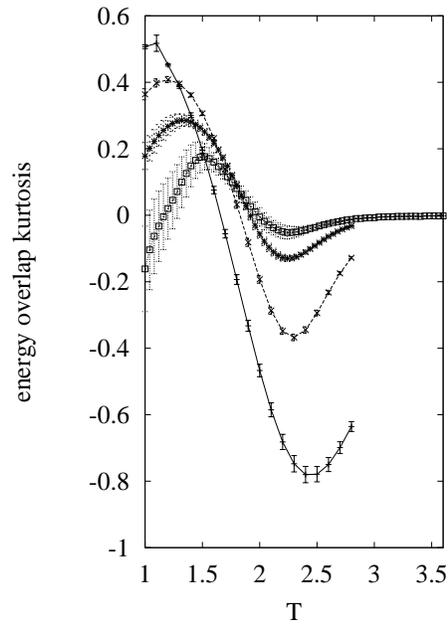}
\end{center}
\caption[0]{\protect\label{F-KUR-E}
  As in figure (\ref{F-CHI}) but for the energy overlap 
  kurtosis $g_{\epsilon}$.}
\end{figure}

Figure (\ref{F-KUR-E}), where we plot the energy overlap kurtosis, is 
dramatically and delightfully different from figure (\ref{F-KUR}).
It is already very interesting to look at the high $T$ region: here 
only the $L=9$ lattice starts to be gaussian, while smaller lattices 
have a strongly non-gaussian behavior. In the cold region again there 
is a strong finite size dependence (we will see in the next section 
that there is a finite size double peak structure that is 
disappearing on large lattices). Here there is a crossing, even if it 
is inverted as compared to usual, $h=0$ systems, where in the warm 
phase the kurtosis curves become lower with increasing lattice size: 
in our case for small sizes and high $T$ we have a negative 
kurtosis, that tends to zero from below when the size increases. This 
real, inverted crossing is at $T$ higher than $1.5$, but we believe it 
should be taken {\em cum grano salis} as far as the exact 
determination of the critical point is concerned. The finite size 
effects make themselves clear in the non-gaussian behavior in the warm 
phase, and in a peak in the cold phase, that shrinks and shifts with 
increasing lattice size (see figure (\ref{F-KUR-E})). A value of 
$T_{c}\simeq 1.5$ is very compatible with this picture.

\begin{figure}
\begin{center}
\leavevmode
\epsfysize=250pt
\epsffile{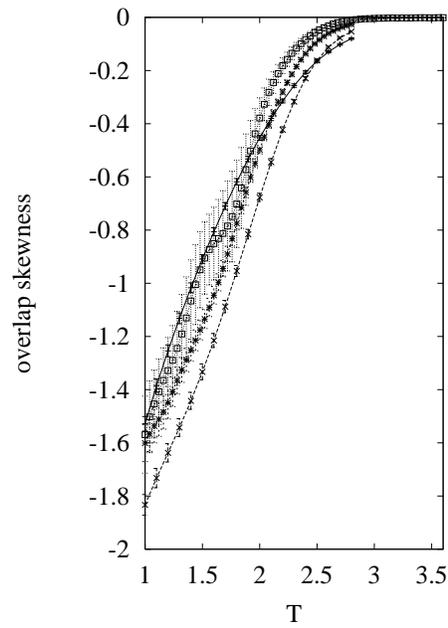}
\end{center}
\caption[0]{\protect\label{F-SKE}
  As in figure (\ref{F-CHI}) but for the overlap skewness $s$.}
\end{figure}

\begin{figure}
\begin{center}
\leavevmode
\epsfysize=250pt
\epsffile{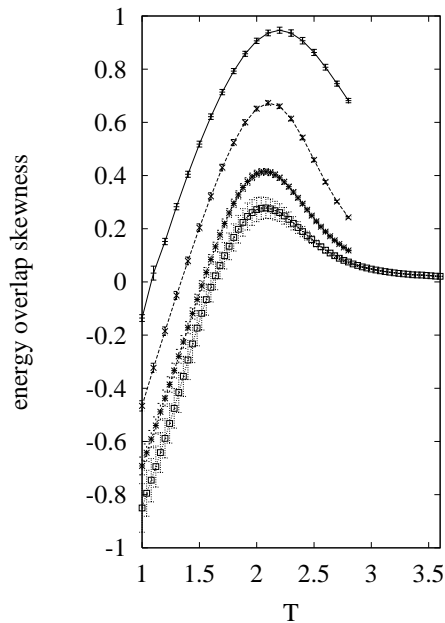}
\end{center}
\caption[0]{\protect\label{F-SKE-E}
  As in figure (\ref{F-CHI}) but for the energy overlap 
  skewness $s_{\epsilon}$.}
\end{figure}

The behavior of the skewness for the overlap (\ref{F-SKE}) and for the 
energy overlap (\ref{F-SKE-E}) repeats a similar pattern.  We find a 
symmetric behavior of the overlap in the warm phase, and curves 
collapse to a non-trivial shape in the low $T$ regime.  The energy 
overlap skewness, on the contrary, starts to become zero in the warm 
region only on our larger lattice, $L=9$, and heavily depends on $L$ 
in the low $T$ region. Again a peak close to $T\simeq 2$ is shrinking
with increasing lattice size.

In the next section we will discuss the full probability distribution.  
The analysis we have discussed here strongly suggests the existence of 
a RSB phase.  We have been able to get hints for the value of $T_{c}$ 
and of the critical exponents, but such estimates, that surely improve 
on existing ones, have to be considered only as hints.

\section{Probability Distributions}

Before starting a detailed discussion of the behavior of the
probability distribution of the order parameter for a given sample
$P_J(q)$ and of the disorder averaged $P(q)$ we briefly 
discuss finite size effects on, for example, $\langle q \rangle_L$. We
have analyzed the size dependence of $\langle q \rangle_L$ and 
$\langle q_E \rangle_L$. We have tried a fit of the type (see
\cite{CPRR} for a detailed discussion of the issue)

\bea
  \langle q \rangle_L &\simeq&  \langle q \rangle_\infty 
   + \frac{A}{L^{d_{q}}}\ ,\\
  \langle q_E \rangle_L &\simeq&  \langle q_E \rangle_\infty 
   + \frac{A_E}{L^{d_{q_E}}}\ .
\eea
In the Parisi solution of the mean field theory $d_q=d_{q_E}$: 
corrections that scale as $V^{-\frac13}$ and the fact that the upper 
critical dimension is $6$ imply that in the mean field limit we 
expect $d_{q}=2$. 

We find that for the overlap the best fit to $d_q$ (using $L=3$, $5$, 
$7$ and $9$) increases with $T$ from $2.3\pm 0.5$ at $T=1.0$ to 
$3.2\pm 0.2$ at $T=1.6$, and to a number larger than $4$ for $T$ among 
$2$ and $3$ (for $T>T_{c}$ we expect an exponential decay, but on a 
finite lattice with a finite number of points we can fit an effective 
exponent).  $d_{q_E}$ is larger: here finite size effects are smaller, 
and the exponent more difficult to determine.  In the region of 
$T=1.6$, $1.8$, $2.0$ we find an exponent close to $4$, that increases 
in the high $T$ region.  For example at $T=1.6$, that asymptotically 
is probably marginally off-critical but very close to the estimated 
$T_{c}$, where we have a clean determination of both exponents, we 
have a ratio $\frac{d_{q_E}}{d_q}\simeq \frac43$ (to be compared with 
the $1$ that one finds in the mean field theory).  The behavior of 
$\langle q \rangle_L$ does not look compatible with a pure 
exponential, while $\langle q_E \rangle_L$ would also be compatible 
with an exponential decay.  We stress again that since the expected 
finite size corrections have a very complex pattern (different terms 
could be leading for different $L$ values) the exact theoretical 
significance of this numerical result is not clear (see \cite{CPRR} 
for a discussion), but for the fact that we find that the two 
exponents are not very different from the ones that are found in mean 
field. For a comparison of the equilibrium value $E(q)$, and the 
dynamical value of the overlap, $q_{D}$, see figure two of ref. 
\cite{MAPAZU} (our data for $E(q)$ are very similar to the ones of 
\cite{PICRIT-B}).

\begin{figure}
\begin{center}
\leavevmode
\epsfysize=250pt
\epsffile{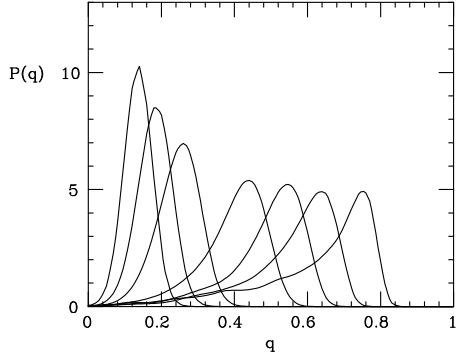}
\end{center}
\caption[0]{\protect\label{F-PQT}
  $P(q)$ at $L=9$ for 
  different $T$ values: from the left are the curves for
  $T=2.4$, $2.2$, $2.0$, 
  $1.6$, $1.4$, $1.2$, $1.0$.}
\end{figure}

Let us examine in some more detail now the full $P(q)$ and $P^{E}(q)$ 
and the probability distributions for a given sample, $P_{J}(q)$ and 
$P^{E}_{J}(q)$.  In figure (\ref{F-PQT}) we plot $P(q)$ at $L=9$ for 
different $T$ values.  At $T=2.4$ $P$ is gaussian and symmetric.  For 
lower $T$ values we get a strong deviation from a Gaussian behavior.  
We will discuss later the fact that finite volume effects are very 
strong (and they turn out to be the most serious limitation in the 
case of the simulations we are discussing here).  We can already 
notice that at $T=1$ from the results of \cite{MAPAZU} we expect that 
the minimum value allowed for the overlap is $q_{min}\simeq 0.55$.  On 
the contrary even on our larger size, $L=9$, we get a long tail that 
basically goes down to $q=0$ (we will see later that it goes down to 
$-1$ for the smaller lattice sizes).

\begin{figure}
\begin{center}
\leavevmode
\epsfysize=250pt
\epsffile{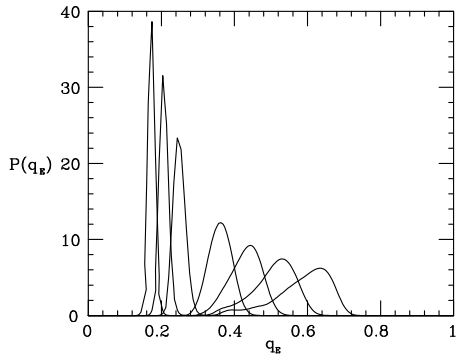}
\end{center}
\caption[0]{\protect\label{F-PQET}
  As in figure (\ref{F-PQT}) but for $P^{E}(q)$.}
\end{figure}

A very similar pattern holds for $P^{E}(q)$ in figure (\ref{F-PQET}). 
Here even at $L=9$ and at $T=2.4$ (deep in the warm phase) the 
probability distribution is not yet fully gaussian (as one can also see 
from the kurtosis and the skewness). As for $P(q)$ the distribution 
becomes very asymmetric at low $T$, with a large tail towards small 
overlaps.

\begin{figure}
\begin{center}
\leavevmode
\epsfysize=250pt
\epsffile{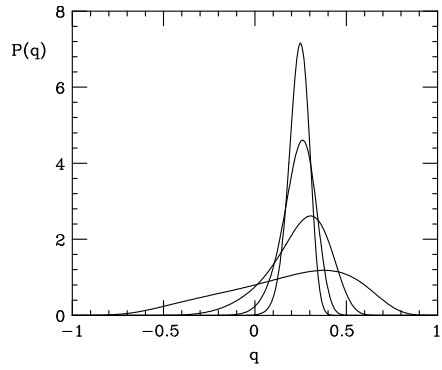}
\end{center}
\caption[0]{\protect\label{F-PQ-TWO-L}
  $P(q)$ versus $q$ at $T=2.0$ for $L=3$, $5$, $7$ and $9$ (curves from 
  bottom to top).}
\end{figure}

It is interesting to analyze in more detail the size dependence of 
the probability distributions. In figure (\ref{F-PQ-TWO-L}) we show 
$P(q)$ versus $q$ at $T=2.0$ for $L=3$, $5$, $7$ and $9$. This is the 
point of the transition in the $h=0$ model, asymptotically in the 
warm region for $h=0.4$. Clearly for $L=3$ the distribution is far 
from Gaussian: only at $L=7$ one sees a Gaussian behavior.

\begin{figure}
\begin{center}
\leavevmode
\epsfysize=250pt
\epsffile{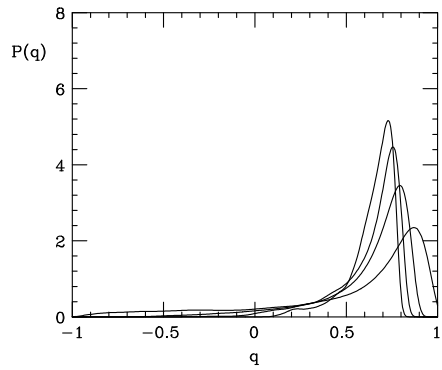}
\end{center}
\caption[0]{\protect\label{F-PQ-ONE-L}
  As in figure (\ref{F-PQ-TWO-L}) but $T=1.0$.}
\end{figure}

The same function at low $T$ is very different.  We show $P(q)$ versus 
$q$ at $T=1.0$ in figure (\ref{F-PQ-ONE-L}).  Also here finite size 
effects are very large, but $P(q)$ does not show any sign of 
convergence to a gaussian behavior.  In the mean field RSB solution 
one would expect a delta function at $q_{min}$ (that, as we already 
said, at $T=1.0$ is close to $0.55$ \cite{MAPAZU}): here, at least for 
$L<9$, we do not see a $\delta$ function like contribution at 
$q<q_{max}$.  Only at $L=9$ we see a small bump at $q\simeq 0.2$: we 
see it in all our different runs, but we cannot exclude (and on the 
contrary we believe it is possible) that it is due to lack of complete 
thermalization of the $L=9$ data at the lower $T$ values (as we have 
discussed even if quantities like $\langle q^{2}\rangle$ are well 
behaved and apparently thermalized we cannot completely exclude a very 
small effect of this kind). Also we will discuss a similar effect, 
that turns out to be due to the finite size of the lattice, for the 
energy overlap.

\begin{figure}
\begin{center}
\leavevmode
\epsfysize=250pt
\epsffile{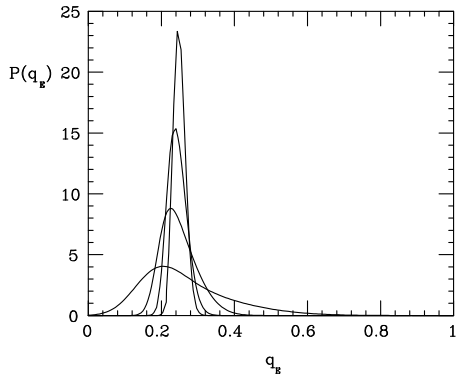}
\end{center}
\caption[0]{\protect\label{F-PQE-TWO-L}
  As in figure (\ref{F-PQ-TWO-L}) but $P_{E}$.}
\end{figure}

In figure (\ref{F-PQE-TWO-L}) we show 
$P_{E}(q_{E})$ versus $q$ at $T=2.0$ for $L=3$, $5$, $7$ and $9$. 
Again, on small lattices even at warm $T$ $P_{E}$ is not symmetric 
(the tail is in this case for large values of the overlap).

\begin{figure}
\begin{center}
\leavevmode
\epsfysize=250pt
\epsffile{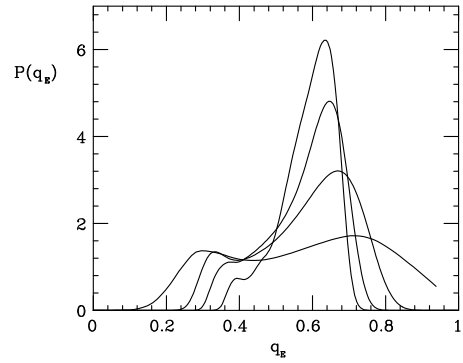}
\end{center}
\caption[0]{\protect\label{F-PQE-ONE-L}
  As in figure (\ref{F-PQ-TWO-L}) but  $P_{E}$ and $T=1.0$.}
\end{figure}

In figure (\ref{F-PQE-ONE-L}) we show that in the cold region ($T=1$) 
the energy overlap has even stronger finite size effects than the 
spin-spin overlap.  Here a spurious peak at low values of $q_{E}$ is 
very strong for small $L$ values (it carries the thirty percent of the 
weight at $L=3$), and becomes smaller and smaller on larger lattices.

\begin{figure}
\begin{center}
\leavevmode
\epsfysize=250pt
\epsffile{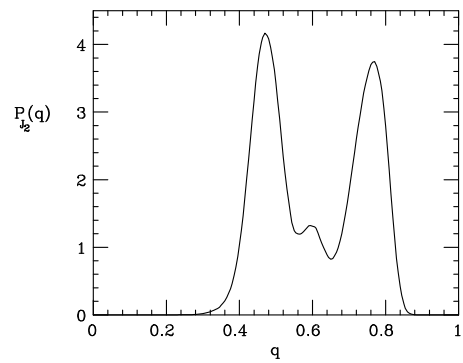}
\end{center}
\caption[0]{\protect\label{F-PQ-S1}
  A first $P_{J}(q)$ for one disorder sample, $L=9$ and $T=1$.}
\end{figure}

\begin{figure}
\begin{center}
\leavevmode
\epsfysize=250pt
\epsffile{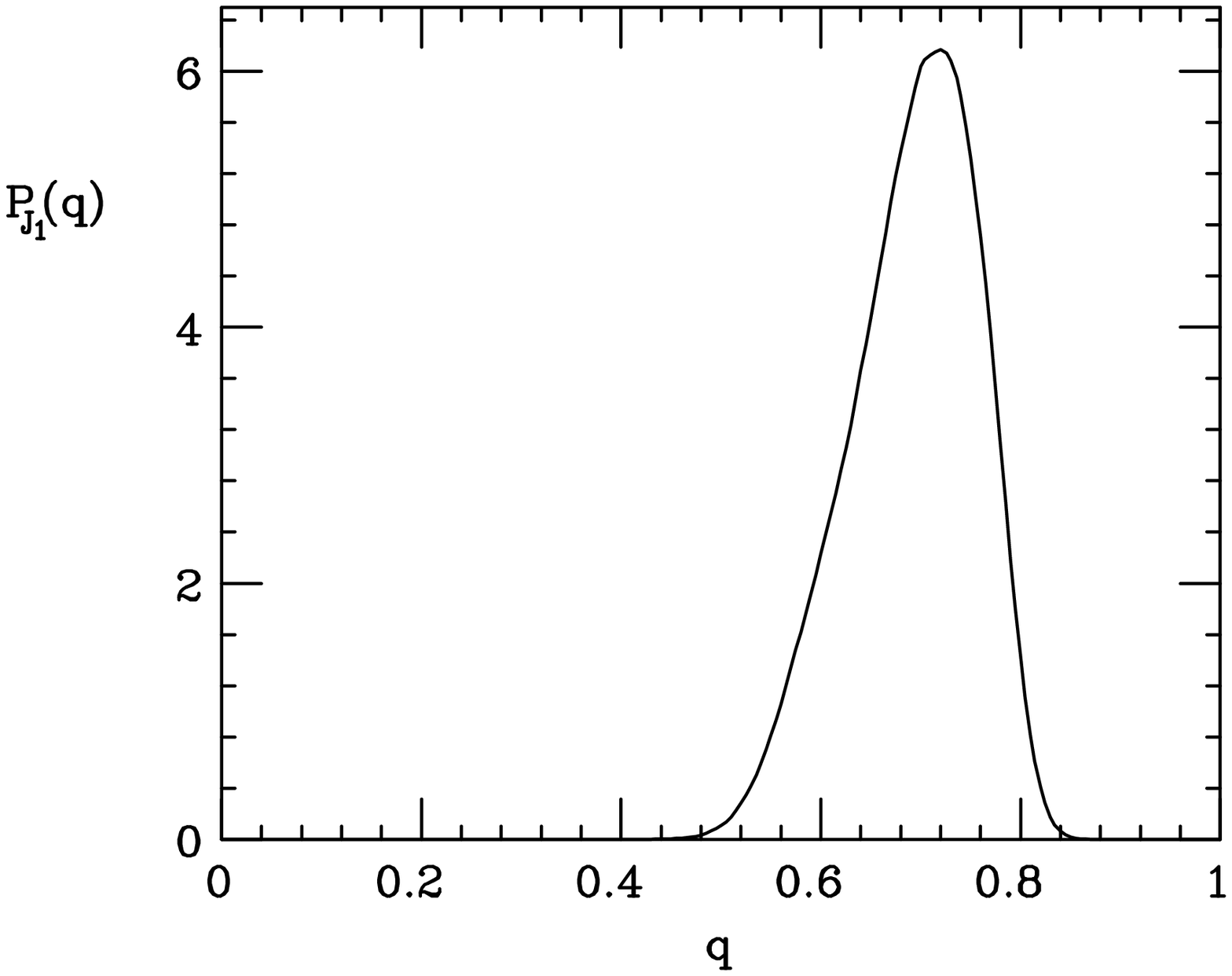}
\end{center}
\caption[0]{\protect\label{F-PQ-S2}
  A second $P_{J}(q)$ for a different disorder sample, $L=9$ and $T=1$.}
\end{figure}

We also show, in figures (\ref{F-PQ-S1}) and (\ref{F-PQ-S2}) two 
individual $P(q)$ for two different realizations of the quenched 
disorder at $L=9$, $T=1$.  It is clear that different samples can have 
a very different equilibrium probability distribution, and the system 
does not look self-averaging.  In the two examples we show, for 
example, it is clear that, like in mean field theory, we can find 
systems with one maximum of $P(q)$ and systems with a complex 
structure of $P(q)$, with many local maxima.  The integrated $P(q)$ is 
not originated from a trivial sum of very similar individual 
$P_{J}(q)$, but from the sum of individual very different 
distributions.  It is clear that this feature, crucial in the RSB mean 
field picture, is shared by the finite dimensional system in a 
non-zero magnetic field.

\section{Conclusions}

We have discussed accurate numerical simulations of the 4D Edwards 
Anderson spin glass in magnetic field.  Our results hint very strongly 
the existence of a spin glass phase transition.  Susceptibilities grow 
strongly at low $T$, and fits to a divergent behavior are very good.  
Cumulants of the overlap and energy overlap probability distribution 
like the kurtosis and the skewness show a clear change of regime in 
the region of temperatures lower than the $h=0$ critical point.  
Finite size effects are dominated by power laws that are similar to 
the ones of the mean field theory.  Probability distributions are 
non-trivial in the low $T$ region, and, what is most important, 
different samples behave clearly in very different ways.  The energy 
overlap follows the usual overlap in this RSB like behavior, making 
the possibility of a non-trivial behavior caused by interfaces quite 
unplausible.

There are also differences with the usual RSB like picture at $h=0$.  
For example here the Binder cumulants do not cross (and the pictures 
of the $P(q)$ show why).  What is more impressive and relevant is the 
presence of very strong finite size effects (even on lattices of size 
$L^{4}=9^{4}$, that had never before been thermalized in a numerical 
simulation).  These effects are far larger than in the $h=0$ case.  
This is the most important limitations of the present numerical 
simulation (and of the physics conclusions one can draw from them): 
when finite size effects are as strong as we have shown is very 
difficult to be sure that any asymptotic behavior has been observed.  
Because of that all the quantitative results we quote for critical 
exponents and temperatures have to be taken as simple indications.
The other real problem, as far as the coincidence with the RSB mean 
field picture is concerned, is that even for $L=9$ we do not see any 
trace of a $\delta$-function at $q=q_{min}$: even if on theoretical 
grounds we expect this peak to be smaller than the one at $q_{max}$ 
\cite{CPRR}, and if we know that finite size effects are very strong, 
and if we have a small bump in $P(q)$ at $L=9$ at low $q$ (that we 
cannot take too seriously), this a worrying point, that is there to 
demand further clarification.

\section*{Acknowledgments}
We thank Giorgio Parisi and Felix Ritort for many helpful conversations.


\end{document}